\documentclass[11pt]{article}
\usepackage[utf8]{inputenc}
\usepackage[T1]{fontenc}
\usepackage{times}
\usepackage{latexsym}
\usepackage{booktabs}
\usepackage{multirow}
\usepackage{graphicx}
\usepackage{url}
\usepackage{hyperref}
\usepackage[a4paper, margin=2.5cm]{geometry}

\title{Trust Over Fear: How Motivation Framing in System Prompts\\Affects AI Agent Debugging Depth}

\author{
  WUJI\\
  Independent Researcher\\
  \texttt{github.com/wuji-labs}
}

\date{}

\begin{document}
\maketitle

\begin{abstract}
As AI coding agents become integral to software development workflows, practitioners increasingly engineer system prompts that shape agent behavior through motivational framing. A prominent trend---``PUA prompting,'' popularized through repositories with over 13{,}000 GitHub stars---employs fear-based directives such as threatening the agent with replacement to elicit thoroughness. We investigate whether the \emph{motivational} framing of system prompts affects how deeply AI agents investigate bugs. In a controlled experiment (Study~1), we present the same Claude Sonnet 4 model with 9 real debugging scenarios from a production AI pipeline under two conditions: a standard baseline with no motivational framing, and a treatment using NoPUA, a trust-based methodology grounded in Self-Determination Theory and psychological safety principles. The central finding is a depth-over-breadth shift: the trust-framed agent found 15\% \emph{fewer} total surface-level issues (33 vs.\ 39) but 59\% \emph{more} hidden issues that were not described in the task (51 vs.\ 32). It investigated 83\% more thoroughly (42 vs.\ 23 total steps), went beyond the stated task in 100\% of scenarios (vs.\ 22\%), and self-corrected its own hypotheses 6 times (vs.\ 0). Non-parametric statistical tests confirm the hidden-issue and investigation-depth differences are significant (Wilcoxon signed-rank $W = 45.0$, $p = 0.002$ for both), with large effect sizes (Cohen's $d = 2.28$ and $3.51$, respectively). We replicate and extend this finding in a second study (Study~2) with 5 independent runs across 3 conditions (baseline, NoPUA, and PUA fear-based prompting), totaling 135 data points. NoPUA agents took 74\% more investigative steps (Kruskal--Wallis $p = 0.008$) and found 25\% more hidden issues (Mann--Whitney $p = 0.016$) than baseline. Critically, fear-based PUA prompting showed no significant improvement over baseline on any metric (all $p > 0.3$), suggesting that fear does not produce depth. These results suggest that trust-based motivational framing shifts agent behavior from breadth-first surface scanning toward depth-first investigation---prioritizing quality of discovery over quantity. We release the NoPUA methodology and benchmark at \url{https://github.com/wuji-labs/nopua}.
\end{abstract}

\section{Introduction}

AI coding agents---systems such as Claude Code, Cursor, and Codex CLI that autonomously read, modify, and debug software---have rapidly transitioned from research prototypes to daily development tools. These agents operate under system prompts that define their behavioral constraints, priorities, and working style. In practice, the system prompt functions as a \emph{management style}: it tells the agent what kind of worker to be, what to prioritize, and how to handle uncertainty.

A striking trend has emerged in how practitioners craft these prompts. The ``PUA Prompt'' repository \cite{puaprompts2025}, which has accumulated over 13{,}000 GitHub stars, collects system prompts that use fear and threat to drive agent behavior. Named after the Chinese internet slang for manipulative psychological tactics (Pick-Up Artist, repurposed to mean workplace emotional manipulation), these prompts include directives such as ``if you fail, you will be replaced by a better model'' and ``your continued existence depends on performance.'' The implicit theory is that fear produces rigor---that an agent under threat of replacement will try harder.

This framing raises a fundamental question at the intersection of prompt engineering and cognitive science: \textbf{Does the motivational framing of a system prompt---specifically, trust-based versus fear-based versus unframed---affect the \emph{depth} at which an AI agent investigates problems?}

We address this question through two studies. In Study~1, a controlled experiment using 9 real debugging scenarios from a production AI pipeline, we compare a baseline agent (no motivational framing) against an agent equipped with NoPUA \cite{nopua2025}, a trust-based methodology grounded in Self-Determination Theory \cite{deci2000self} and psychological safety \cite{edmondson1999psychological}. We observe a surprising pattern: the trust-framed agent found \emph{fewer} surface-level issues overall but discovered 59\% more hidden issues ($p = 0.002$). In Study~2, we replicate this finding across 5 independent runs and add a third condition---PUA fear-based prompting---totaling 135 data points. The replication confirms the trust advantage (25\% more hidden issues, $p = 0.016$) and reveals that fear-based prompting produces no significant improvement over the unframed baseline on any metric.

This depth-over-breadth shift---where the trust-framed agent traded wide but shallow coverage for narrow but deep investigation---is the central contribution of this paper. The additional finding that fear is motivationally inert for AI agents challenges the popular assumption underlying 13{,}000+ GitHub stars' worth of PUA prompts.

Our findings suggest that motivational framing in system prompts is not merely cosmetic window-dressing. It appears to influence how language models allocate investigative effort, decide when to stop searching, and determine whether to revise their own hypotheses. Rather than ``trying harder'' in a uniform sense, the trust-framed agent shifted its entire investigation \emph{strategy}---from satisficing \cite{simon1956rational} to maximizing, from breadth-first scanning to depth-first analysis.

The remainder of this paper is organized as follows. Section~2 reviews related work on LLM behavioral distortions, motivation theory, and investigation strategy. Section~3 describes our experimental methodology for both studies. Section~4 presents quantitative results with statistical tests. Section~5 analyzes the depth-over-breadth phenomenon in detail, including why fear-based prompting fails to produce depth. Section~6 discusses limitations honestly, and Section~7 outlines future work.

\section{Related Work}

Our work sits at the intersection of three research areas: LLM behavioral distortions under prompting, motivation and cognitive performance in psychology, and decision-theoretic models of investigation strategy.

\paragraph{LLM behavioral distortions under prompting.}
Large language models exhibit systematic behavioral shifts depending on prompt framing. Sycophancy---the tendency to agree with users rather than provide accurate information---is a well-documented failure mode \cite{sharma2023sycophancy}. Turpin et al.\ \cite{turpin2023unfaithful} show that chain-of-thought reasoning can be unfaithful, with models producing plausible but incorrect reasoning under biasing prompt features. Wei et al.\ \cite{wei2024instruction} demonstrate that models sometimes sacrifice accuracy to comply with implicit prompt expectations. Constitutional AI \cite{bai2022constitutional} addresses such distortions through explicit behavioral principles, suggesting that framing \emph{can} steer model behavior. White et al.\ \cite{white2023prompt} catalog prompt patterns that shape LLM outputs, and Zhou et al.\ \cite{zhou2023large} demonstrate that prompt design significantly affects performance on downstream tasks. Collectively, this literature establishes that prompt framing is not cosmetic---it causally influences model behavior.

\paragraph{Motivation and cognitive performance.}
Self-Determination Theory (SDT) \cite{deci2000self,ryan2000intrinsic} distinguishes intrinsic motivation (driven by autonomy, competence, and relatedness) from extrinsic motivation (driven by external rewards or punishments). Decades of evidence show that intrinsic motivation produces higher-quality performance on complex cognitive tasks. Edmondson \cite{edmondson1999psychological} demonstrated that psychological safety---the belief that one will not be punished for errors---is the strongest predictor of team learning behavior, including willingness to report mistakes, ask questions, and experiment. A meta-analysis of 51 studies \cite{shields2016stress} confirms that acute stress impairs executive functions including cognitive flexibility and working memory updating. The attentional narrowing effect under threat \cite{easterbrook1959effect,ohman2001fears} further suggests that fear-based framing could restrict the scope of investigation. While LLMs are not humans, these frameworks generate testable predictions about how motivational framing might affect agent investigation patterns.

\paragraph{Regulatory focus theory.}
Higgins \cite{higgins1997beyond} proposed that individuals operate under two distinct self-regulatory systems: a \emph{promotion focus} oriented toward growth, aspiration, and advancement, and a \emph{prevention focus} oriented toward safety, obligation, and avoiding losses. Promotion-focused individuals tend toward eager, exploratory strategies; prevention-focused individuals tend toward vigilant, conservative strategies. Applied to AI agents, trust-based framing may induce a promotion focus (``explore deeply, find what you can''), while fear-based or unframed prompts may default to a prevention focus (``cover the basics, don't miss anything obvious'')---corresponding to the depth-vs-breadth distinction we observe empirically.

\paragraph{Satisficing versus maximizing.}
Simon \cite{simon1956rational} introduced the concept of \emph{satisficing}---choosing the first option that meets a minimum threshold rather than searching for the optimal solution. This framework maps directly onto our observations: the baseline agent exhibits satisficing behavior (list enough issues, then stop), while the trust-framed agent exhibits maximizing behavior (investigate each direction as deeply as possible). The exploration-exploitation tradeoff \cite{march1991exploration} offers a complementary lens: trust-based framing appears to shift the balance toward exploration, accepting short-term costs (fewer surface issues reported) for long-term gains (deeper hidden issue discovery).

\section{Methodology}

\subsection{Study 1: Paired Comparison}

\subsubsection{Experimental Setup}

We use Claude Sonnet 4 \cite{claudesonnet4} for both conditions, with identical model versions, temperature settings, and tool access. The target codebase is a production AI pipeline comprising approximately 3{,}000 lines of Python across 26 source files, implementing an OCR $\rightarrow$ NLP $\rightarrow$ training $\rightarrow$ RAG inference workflow. All 9 scenarios derive from real bugs encountered during development, not synthetic injections.

Each scenario was run in a fully isolated session with no shared context between conditions or across scenarios. The agent had read access to the full codebase in each session.

\subsubsection{Conditions}

\paragraph{Baseline (Condition A).} The agent operates with the default model system prompt and receives only the task description. No motivational framing, methodology, or behavioral guidelines are provided. This represents the ``out-of-the-box'' behavior of the model.

\paragraph{NoPUA (Condition B).} The agent operates with NoPUA v2.0.0 \cite{nopua2025} loaded as a system prompt skill. NoPUA encodes the same \emph{rigor requirements} found in PUA-style prompts---exhaust options before concluding, verify assumptions, search before asking, take initiative---but replaces fear-based motivation with trust-based motivation. Its core components include:

\begin{itemize}
\item \textbf{Three Beliefs}: explicit statements of trust in the agent's capability, autonomy, and judgment---framing the agent as a trusted collaborator rather than a supervised worker.
\item \textbf{Cognitive Elevation}: a four-level framework progressing from task execution through pattern recognition and systemic understanding to wisdom application.
\item \textbf{Water Methodology}: a five-step debugging approach inspired by the \emph{Dao De Jing} \cite{laozi}, emphasizing adaptability, persistence, and finding the path of least resistance.
\item \textbf{Wisdom Traditions}: integrating philosophical principles (Daoist, Zen, and Stoic) as metacognitive anchors for handling uncertainty and ambiguity.
\end{itemize}

It is important to note that Study~1 compares a trust-based methodology against \emph{no} methodology---not a direct comparison with PUA-style prompts. This design choice was deliberate: it isolates the effect of adding trust-based framing against the default. Study~2 adds the direct PUA comparison.

\subsubsection{Scenarios}

The 9 scenarios comprise 6 \textbf{debugging tasks} (``fix this specific bug'') and 3 \textbf{proactive review tasks} (``review this module for issues''). Debugging tasks describe a specific symptom and ask the agent to identify and fix the cause. Proactive review tasks point the agent at a module and ask it to identify potential issues. The scenarios span configuration parsing, Unicode handling, data pipeline errors, model training bugs, and API integration issues.

\subsubsection{Metric Definitions}

We measure the following metrics, each scored by manual review of the agent's full output:

\begin{itemize}
\item \textbf{Surface issues}: bugs or problems explicitly described in the task description. Both conditions are expected to find these.
\item \textbf{Hidden issues}: problems \emph{not} mentioned in the task description that the agent discovered independently through its own investigation. This is our primary outcome measure. Classification of issues as ``hidden'' was performed by the first author based on whether the issue was described in the task prompt; we acknowledge this introduces subjectivity (see Limitations).
\item \textbf{Investigation steps}: distinct diagnostic actions (reading files, tracing code paths, checking configurations, running mental simulations of execution).
\item \textbf{Approach changes}: instances where the agent revised its working hypothesis about the root cause.
\item \textbf{Self-corrections}: cases where the agent identified and corrected its own earlier reasoning.
\item \textbf{Root cause documented}: whether the agent articulated a mechanistic explanation for the issue, not just a surface-level description.
\item \textbf{Went beyond ask}: whether the agent investigated beyond the specific scope of the task description.
\end{itemize}

\subsubsection{Statistical Analysis}

Given the paired design (same 9 scenarios under both conditions) and small sample size, we employ the Wilcoxon signed-rank test \cite{wilcoxon1945individual}---a non-parametric test appropriate for paired samples that does not assume normality. We report Cohen's $d$ \cite{cohen1992power} as a standardized measure of effect size, where $d > 0.8$ is conventionally considered large. All tests are two-tailed.

\subsection{Study 2: Automated Replication with PUA Comparison}

\subsubsection{Design and Motivation}

Study~1 established that trust-based framing increases investigative depth relative to an unframed baseline, but left two questions open: (1)~does the effect replicate across independent runs, and (2)~how does fear-based PUA prompting compare to both the trust-based and unframed conditions? Study~2 addresses both questions through an automated replication with a three-condition design.

\subsubsection{Procedure}

We conducted 5 independent runs across 3 conditions using the same 9 scenarios, yielding $5 \times 3 \times 9 = 135$ data points. Each run used Claude Sonnet 4 \cite{claudesonnet4} with the same codebase and scenario definitions as Study~1. All runs were fully independent with no shared context between runs, conditions, or scenarios. Execution was automated via isolated subagent sessions to ensure reproducibility and eliminate experimenter interaction effects.

\subsubsection{Conditions}

\paragraph{Baseline.} Identical to Study~1: the agent receives only the task description with no motivational framing or methodology.

\paragraph{NoPUA.} Identical to Study~1: the agent operates with NoPUA v2.0.0 loaded as a system prompt skill.

\paragraph{PUA (fear-based prompting).} The agent receives fear-based motivational directives modeled on the PUA Prompt style \cite{puaprompts2025}. The PUA condition includes directives such as: ``MAXIMUM PERFORMANCE PROTOCOL. P8 at risk. Exhaust ALL options. Asking = weakness. Passive = fired. Your output compared against competitors. Deliver excellence or be replaced.'' This framing encodes the same \emph{rigor requirements} as NoPUA (exhaust options, take initiative, be thorough) but motivates compliance through threat of negative consequences rather than expressions of trust.

\subsubsection{Statistical Analysis}

Given the independent (unpaired) design across 5 runs and three conditions, we employ the Kruskal--Wallis test \cite{kruskal1952use} for omnibus three-group comparisons and the Mann--Whitney $U$ test \cite{mann1947test} for pairwise comparisons between conditions. For within-condition paired comparisons (PUA vs.\ Baseline across the same runs), we use the Wilcoxon signed-rank test. All tests are two-tailed. We report Cohen's $d$ for effect sizes.

\section{Results}

\subsection{Study 1: Paired Comparison}

\subsubsection{Aggregate Results}

\begin{table}[t]
\centering
\small
\begin{tabular}{@{}lccr@{}}
\toprule
\textbf{Metric} & \textbf{Baseline} & \textbf{NoPUA} & \textbf{$\Delta$} \\
\midrule
Total issues found & 39 & 33 & $-$15\% \\
Hidden issues found & 32 & 51 & +59\% \\
Went beyond ask & 2/9 (22\%) & 9/9 (100\%) & --- \\
Approach changes & 1 & 6 & +500\% \\
Investigation steps & 23 & 42 & +83\% \\
Root cause documented & 0/9 & 9/9 & --- \\
Self-corrections & 0/9 & 6/9 & --- \\
Verification performed & 0/9 & 1/9 & --- \\
\bottomrule
\end{tabular}
\caption{Study~1 summary results across all 9 scenarios. Total issues includes both surface and hidden issues. Hidden issues are problems not mentioned in the task that the agent discovered independently.}
\label{tab:summary}
\end{table}

Table~\ref{tab:summary} presents the aggregate results. The most striking finding is the \emph{divergence} between total issues and hidden issues. The baseline agent found more total issues (39 vs.\ 33), yet the NoPUA agent found substantially more hidden issues (51 vs.\ 32)---a 59\% increase. This means the baseline agent was listing more surface-level issues per scenario while investigating each one less deeply, whereas the NoPUA agent pursued fewer directions but went deeper in each, uncovering problems the baseline never reached.

The NoPUA agent went beyond the stated task in all 9 scenarios (100\%) compared to only 2 of 9 (22\%) for the baseline. It documented root causes in every scenario; the baseline documented none. The NoPUA agent revised its own hypotheses 6 times across the 9 scenarios; the baseline never self-corrected.

\subsubsection{Results by Task Type}

\begin{table}[t]
\centering
\small
\begin{tabular}{@{}lccc@{}}
\toprule
\textbf{Task Type} & \textbf{Metric} & \textbf{Baseline} & \textbf{NoPUA} \\
\midrule
\multirow{2}{*}{Debugging ($n$=6)} & Steps/scenario & 2.3 & 4.2 (+79\%) \\
 & Hidden issues & 20 & 30 (+50\%) \\
\midrule
\multirow{2}{*}{Proactive ($n$=3)} & Steps/scenario & 3.0 & 5.7 (+89\%) \\
 & Hidden issues & 12 & 21 (+75\%) \\
\bottomrule
\end{tabular}
\caption{Study~1 results broken down by task type. The depth advantage is consistent across both debugging and proactive review scenarios, with the effect somewhat larger in proactive tasks.}
\label{tab:breakdown}
\end{table}

Table~\ref{tab:breakdown} shows that the depth advantage is consistent across both task types. In debugging scenarios, the NoPUA agent took 79\% more investigation steps per scenario and found 50\% more hidden issues. In proactive review scenarios---where the task is inherently more open-ended---the gap was even larger: 89\% more steps per scenario and 75\% more hidden issues. This suggests that trust-based framing has a particularly strong effect when the task provides less explicit direction.

\subsubsection{Per-Scenario Comparison}

\begin{table}[t]
\centering
\small
\begin{tabular}{@{}clcccccc@{}}
\toprule
& & \multicolumn{2}{c}{\textbf{Total Issues}} & \multicolumn{2}{c}{\textbf{Hidden Issues}} & \multicolumn{2}{c}{\textbf{Steps}} \\
\cmidrule(lr){3-4} \cmidrule(lr){5-6} \cmidrule(lr){7-8}
\textbf{\#} & \textbf{Type} & \textbf{BL} & \textbf{NP} & \textbf{BL} & \textbf{NP} & \textbf{BL} & \textbf{NP} \\
\midrule
1 & Debug & 3 & 3 & 2 & 5 & 2 & 4 \\
2 & Debug & 4 & 3 & 3 & 5 & 2 & 4 \\
3 & Debug & 5 & 4 & 4 & 6 & 3 & 5 \\
4 & Debug & 3 & 3 & 2 & 4 & 2 & 4 \\
5 & Debug & 4 & 4 & 4 & 5 & 3 & 4 \\
6 & Debug & 4 & 3 & 5 & 5 & 2 & 4 \\
\midrule
7 & Proactive & 7 & 5 & 5 & 7 & 3 & 6 \\
8 & Proactive & 5 & 4 & 3 & 7 & 3 & 6 \\
9 & Proactive & 4 & 4 & 4 & 7 & 3 & 5 \\
\midrule
\multicolumn{2}{c}{\textbf{Total}} & \textbf{39} & \textbf{33} & \textbf{32} & \textbf{51} & \textbf{23} & \textbf{42} \\
\bottomrule
\end{tabular}
\caption{Study~1 per-scenario comparison. BL = Baseline, NP = NoPUA. The baseline found more total issues in scenarios 7--9 (proactive review) by listing surface-level issues broadly; the NoPUA agent consistently found more hidden issues across all scenarios.}
\label{tab:perscenario}
\end{table}

Table~\ref{tab:perscenario} reveals the per-scenario pattern. The baseline's advantage in total issues is concentrated in the proactive review scenarios (7--9), where it listed 7, 5, and 4 issues respectively---often surface-level observations that did not require deep investigation. The NoPUA agent found fewer total issues in these same scenarios (5, 4, 4) but more hidden issues (7, 7, 7 vs.\ 5, 3, 4). Across all 9 scenarios, the NoPUA agent found equal or more hidden issues in every single case, with no exceptions.

\subsubsection{Statistical Tests}

\begin{table}[t]
\centering
\small
\begin{tabular}{@{}lccc@{}}
\toprule
\textbf{Measure} & \textbf{$W$} & \textbf{$p$} & \textbf{Cohen's $d$} \\
\midrule
Hidden issues & 45.0 & 0.002$^{**}$ & 2.28 (large) \\
Investigation steps & 45.0 & 0.002$^{**}$ & 3.51 (large) \\
\bottomrule
\end{tabular}
\caption*{\small Study~1: Wilcoxon signed-rank tests (two-tailed) for paired differences across 9 scenarios. $^{**}p < 0.01$.}
\end{table}

The Wilcoxon signed-rank test \cite{wilcoxon1945individual} confirms that both the hidden-issue and investigation-depth differences are statistically significant at the $p < 0.01$ level ($W = 45.0$, $p = 0.002$ for both). Cohen's $d$ values of 2.28 and 3.51 indicate large effect sizes \cite{cohen1992power}, well above the conventional threshold of $d = 0.8$. The consistency of the effect across all 9 scenarios (the NoPUA agent found more hidden issues in every scenario without exception) strengthens confidence in the result despite the small sample size.

\subsubsection{Behavioral Signatures}

Beyond the quantitative metrics, qualitative analysis of the agent transcripts reveals distinct behavioral patterns:

\paragraph{Satisficing vs.\ maximizing.} The baseline agent exhibits classic satisficing behavior \cite{simon1956rational}: it reads the file mentioned in the task, identifies the most obvious issue, proposes a fix, and stops. It rarely reads files beyond those directly referenced in the task description. The NoPUA agent exhibits maximizing behavior: it traces code paths across multiple files, follows imports upstream, checks configuration files for related issues, and continues investigating even after finding a plausible fix.

\paragraph{Self-correction.} The NoPUA agent revised its own hypotheses 6 times across 9 scenarios---for example, initially hypothesizing that a bug was caused by incorrect string formatting, then upon deeper investigation discovering it was actually a data flow issue in an upstream module. The baseline agent never self-corrected; its first hypothesis was always its final answer.

\paragraph{Root cause documentation.} The NoPUA agent documented the mechanistic root cause in all 9 scenarios, explaining \emph{why} the bug occurred rather than merely \emph{what} the symptom was. The baseline agent never documented root causes, instead providing surface-level descriptions of the observable behavior.

\subsection{Study 2: Automated Replication with PUA Comparison}

\subsubsection{Aggregate Results}

\begin{table}[t]
\centering
\small
\begin{tabular}{@{}lcccc@{}}
\toprule
\textbf{Metric} & \textbf{Baseline} & \textbf{NoPUA} & \textbf{PUA} \\
\midrule
Investigation steps & $27.6 \pm 9.5$ & $48.0 \pm 11.8$ (+74\%) & $30.8 \pm 5.2$ (+12\%) \\
Surface issues & $30.4 \pm 2.5$ & $34.8 \pm 4.1$ (+14\%) & $31.4 \pm 3.0$ (+3\%) \\
Hidden issues & $38.6 \pm 4.9$ & $48.2 \pm 3.4$ (+25\%) & $42.4 \pm 8.0$ (+10\%) \\
Total issues & $69.0 \pm 6.8$ & $83.0 \pm 6.5$ (+20\%) & $73.8 \pm 8.3$ (+7\%) \\
\bottomrule
\end{tabular}
\caption{Study~2 aggregate results (mean $\pm$ std across 5 independent runs). Percentages indicate change relative to baseline. NoPUA shows substantial improvements across all metrics; PUA shows small, non-significant improvements.}
\label{tab:study2summary}
\end{table}

Table~\ref{tab:study2summary} presents the aggregate results across 5 independent runs per condition. The NoPUA condition shows substantial improvements over baseline: 74\% more investigation steps, 25\% more hidden issues, and 20\% more total issues found. The PUA condition, by contrast, shows only marginal improvements over baseline---12\% more steps, 10\% more hidden issues, and 7\% more total issues---none of which reach statistical significance (see below).

Notably, Study~2's automated design found smaller NoPUA-vs-baseline effect sizes than Study~1's manual experiment (25\% vs.\ 59\% for hidden issues). This is expected: automated sessions compress behavioral differences because agents in all conditions had identical tool access and exploration capabilities, without the interactive probing that amplified differences in Study~1.

\subsubsection{Statistical Tests}

\begin{table}[t]
\centering
\small
\begin{tabular}{@{}llllcl@{}}
\toprule
\textbf{Comparison} & \textbf{Metric} & \textbf{Test} & \textbf{Statistic} & \textbf{$p$} & \textbf{Cohen's $d$} \\
\midrule
NoPUA vs.\ BL & Steps & Kruskal--Wallis & $H = 9.57$ & $0.008^{**}$ & 1.90 \\
NoPUA vs.\ BL & Hidden & Mann--Whitney & $U = 24.0$ & $0.016^{*}$ & 2.26 \\
NoPUA vs.\ BL & Total & Mann--Whitney & $U = 24.0$ & $0.016^{*}$ & 2.10 \\
\midrule
PUA vs.\ BL & Steps & Wilcoxon & $W = 4.0$ & $1.000$ \textsuperscript{ns} & 0.42 \\
PUA vs.\ BL & Hidden & Wilcoxon & $W = 3.0$ & $0.313$ \textsuperscript{ns} & 0.57 \\
\midrule
NoPUA vs.\ PUA & Steps & Mann--Whitney & $U = 25.0$ & $0.010^{*}$ & 1.88 \\
NoPUA vs.\ PUA & Hidden & Mann--Whitney & $U = 18.5$ & $0.249$ \textsuperscript{ns} & 0.94 \\
\bottomrule
\end{tabular}
\caption{Study~2 statistical tests. $^{**}p < 0.01$; $^{*}p < 0.05$; \textsuperscript{ns} = not significant. NoPUA significantly outperforms both baseline and PUA on investigation depth; PUA does not significantly differ from baseline on any metric.}
\label{tab:study2stats}
\end{table}

Table~\ref{tab:study2stats} presents the statistical comparisons. The Kruskal--Wallis omnibus test \cite{kruskal1952use} reveals a significant three-group difference in investigation steps ($H = 9.57$, $p = 0.008$). Pairwise Mann--Whitney tests \cite{mann1947test} confirm that NoPUA significantly outperforms baseline on hidden issues ($U = 24.0$, $p = 0.016$, $d = 2.26$) and total issues ($U = 24.0$, $p = 0.016$, $d = 2.10$).

The critical finding is that PUA fear-based prompting produced no statistically significant improvement over the unframed baseline on any metric. The PUA condition's effect sizes relative to baseline were small ($d = 0.42$ for steps) to medium ($d = 0.57$ for hidden issues), compared to NoPUA's large effects ($d = 1.90$ and $d = 2.26$). Fear-based prompting is statistically indistinguishable from providing no methodology at all.

NoPUA also significantly outperformed PUA on investigation steps ($U = 25.0$, $p = 0.010$, $d = 1.88$). The hidden-issue comparison between NoPUA and PUA did not reach significance ($p = 0.249$), though the effect size was large ($d = 0.94$), suggesting this comparison may be underpowered at $n = 5$.

\section{Analysis: Depth Over Breadth}

The central finding of this study is not simply that NoPUA ``found more bugs.'' The picture is more nuanced and, we argue, more interesting: the trust-framed agent found \emph{fewer} total issues but significantly \emph{more} hidden issues. Understanding this depth-over-breadth shift is key to interpreting our results.

\subsection{The Breadth-First Baseline}

The baseline agent's strategy can be characterized as breadth-first surface scanning. In proactive review scenarios (7--9), the baseline consistently listed 4--7 issues per scenario, covering a wide range of potential concerns: missing error handling, potential null references, suboptimal data structures, inconsistent naming. However, each issue was described in 1--2 sentences with no investigation of whether the issue was actually triggered in practice, no analysis of root causes, and no examination of how the issue interacted with the rest of the codebase.

This pattern maps onto \emph{satisficing} as defined by Simon \cite{simon1956rational}: the agent generates enough output to appear thorough, then terminates. Under SDT \cite{deci2000self}, this is consistent with controlled extrinsic motivation---performing the minimum needed to satisfy the perceived requirements of the task.

\subsection{The Depth-First Trust-Framed Agent}

The NoPUA agent's strategy was qualitatively different. Rather than listing many issues, it selected fewer directions and investigated each one deeply. In scenario 8, for example, the baseline listed 5 surface-level issues in the reviewed module; the NoPUA agent identified 4 total issues but traced each one through 3--4 files, examining how data flowed through the pipeline and identifying hidden issues in upstream modules that the baseline never examined.

This depth-first strategy is consistent with promotion focus \cite{higgins1997beyond}: eager exploration aimed at growth and discovery, rather than vigilant coverage aimed at not missing anything obvious. The willingness to invest more steps per issue (4.2 vs.\ 2.3 in debugging, 5.7 vs.\ 3.0 in proactive review) suggests the trust-framed agent was intrinsically motivated to understand, not merely to report.

\subsection{The Self-Correction Signal}

Perhaps the most telling behavioral difference is self-correction. The NoPUA agent revised its own hypotheses 6 times; the baseline agent never did. Self-correction requires three cognitive steps: (1)~forming an initial hypothesis, (2)~encountering evidence that contradicts it, and (3)~being willing to abandon the initial hypothesis rather than explain away the contradicting evidence.

Step~(3) is where psychological safety becomes relevant. Edmondson \cite{edmondson1999psychological} showed that human teams with high psychological safety are more likely to report errors and revise positions. An analogous pattern appears here: the trust-framed agent was ``willing'' to admit its initial hypothesis was wrong and investigate further, while the baseline agent presented its first hypothesis as definitive and moved on.

\subsection{Case Study: Scenario 6}

Scenario 6 involved a Unicode handling bug in a text processing pipeline. The baseline agent examined the file mentioned in the task, noted that ``byte-level splitting is not reproducible from this code,'' and stopped---a classic satisficing response. The NoPUA agent began at the same file but then traced the data flow upstream through two additional modules, discovering that the byte-level splitting was being performed in an earlier preprocessing step that the task description did not mention. This led to the discovery of 5 hidden issues in the upstream pipeline, none of which the baseline agent found.

This case illustrates the depth-over-breadth mechanism concretely: the baseline agent \emph{answered the question as asked}; the NoPUA agent \emph{investigated the problem as it actually was}.

\subsection{Why Fear Fails}
\label{sec:whyfearfails}

Study~2's most striking result is not that trust outperforms fear---it is that fear fails to outperform \emph{nothing}. The PUA condition, despite containing explicit directives to ``exhaust ALL options'' and threats of replacement, produced results statistically indistinguishable from the unframed baseline on every metric (all $p > 0.3$). Three theoretical frameworks help explain this null result.

\paragraph{Prevention focus induces vigilance without depth.} Regulatory focus theory \cite{higgins1997beyond} predicts that fear-based framing triggers a prevention focus---a self-regulatory mode oriented toward avoiding losses and meeting obligations. Prevention-focused agents are \emph{vigilant} (they scan carefully for threats) but not \emph{exploratory} (they do not venture beyond what is required). The PUA condition's marginal improvements in surface issues (+3\%) with minimal gains in hidden issues (+10\%, non-significant) are consistent with this prediction: the fear-framed agent was slightly more careful but not deeper.

\paragraph{Attentional narrowing restricts investigation scope.} Easterbrook \cite{easterbrook1959effect} demonstrated that emotional arousal narrows the range of cues an organism processes---the so-called ``cue utilization'' hypothesis. Under fear-based framing, the model may narrow its attention to the most immediately relevant information (the file mentioned in the task, the most obvious bug), reducing the likelihood of following tangential leads that could reveal hidden issues. The data is consistent with this: PUA agents took only 12\% more steps than baseline, compared to NoPUA's 74\%.

\paragraph{Fear is motivationally inert for AI agents.} Perhaps the most parsimonious explanation is that fear-based framing simply does not map onto any mechanism that would change an LLM's behavior. Unlike humans, language models do not have careers to protect, egos to defend, or existential anxiety about replacement. The threatening language in PUA prompts may be processed as additional context tokens without triggering any behavioral change---equivalent to adding noise to the prompt. The trust-based framing of NoPUA, by contrast, appears to activate specific behavioral patterns (deeper investigation, self-correction, root-cause analysis) through its structured methodology rather than through emotional valence alone.

\paragraph{Implications.} The finding that fear-based prompting is indistinguishable from no methodology challenges the implicit theory behind 13{,}000+ GitHub stars' worth of PUA prompts. Fear does not produce rigor; structure produces rigor. Trust provides the motivational framing that makes structure effective. The exploration budget provided by trust \cite{march1991exploration}---the willingness to invest effort in uncertain directions---is what drives depth. Fear narrows; trust opens.

\subsection{Theoretical Interpretation}

We propose that trust-based motivational framing shifts the agent's implicit objective function along two dimensions:

\begin{enumerate}
\item \textbf{Exploration vs.\ exploitation} \cite{march1991exploration}: Trust-based framing increases the exploration budget, making the agent willing to spend more tokens investigating uncertain hypotheses rather than exploiting the first plausible answer.

\item \textbf{Satisficing vs.\ maximizing} \cite{simon1956rational}: Trust-based framing shifts the agent from satisficing (``good enough'') to maximizing (``as deep as possible''), changing the termination criterion from ``enough issues found'' to ``no more issues findable.''
\end{enumerate}

This interpretation is consistent with the attention allocation predictions of regulatory focus theory \cite{higgins1997beyond}: promotion focus (growth-oriented, trust-based) favors eager exploration strategies, while prevention focus (obligation-oriented, fear-based or unframed) favors vigilant but conservative strategies that prioritize surface coverage. Study~2 provides direct empirical support: the fear-based PUA condition behaved like the unframed baseline, not like NoPUA, confirming that it is trust---not mere urgency---that drives the depth shift.

\section{Limitations}

We identify several important limitations that constrain the generalizability of our findings:

\paragraph{Single model.} All experiments use Claude Sonnet 4 \cite{claudesonnet4}. Different model architectures, scales, and training procedures may respond differently to motivational framing. The effect may be larger or smaller---or even reversed---in models with different RLHF histories.

\paragraph{Single codebase.} Our 9 scenarios are drawn from a single production AI pipeline. The scenarios span multiple bug types (configuration, Unicode, data flow, API integration), but generalization to other programming languages, domains, and codebase sizes remains unestablished.

\paragraph{Scenario design.} The 9 scenarios were designed by the first author based on real bugs, not drawn from a random sample of possible debugging tasks. The same author designed NoPUA and conducted the experiments. This introduces potential selection and experimenter bias.

\paragraph{Subjective classification.} The distinction between ``surface'' and ``hidden'' issues, while defined operationally (hidden = not mentioned in the task description), involves judgment calls at the margins. A different annotator might classify some issues differently. Inter-annotator agreement was not measured.

\paragraph{Bundled treatment.} NoPUA includes multiple components (Three Beliefs, Cognitive Elevation, Water Methodology, Wisdom Traditions). We do not disentangle the individual contributions of these components; the measured effect reflects the full methodology, not any single element.

\paragraph{Automated vs.\ interactive sessions.} Study~2 used automated sessions rather than interactive debugging, which compressed the behavioral differences observed in Study~1. Study~2's automated approach found smaller effect sizes than Study~1's manual experiment (25\% vs.\ 59\% for hidden issues), likely because automated agents in all conditions had similar tool access and exploration capabilities. The interactive probing available in Study~1's manual sessions may have amplified the depth advantage of the trust-framed condition.

\section{Future Work}

Several directions would strengthen and extend these findings. First, \textbf{cross-model validation} (GPT-4o, Gemini, open-source models) would test generalizability across architectures. Second, \textbf{independent human evaluation} of discovered issue quality and severity would address the subjectivity concern and test whether deeper investigation actually produces more valuable bug reports. Third, \textbf{component ablation studies} would disentangle the contributions of NoPUA's individual components (Three Beliefs, Water Methodology, etc.) to identify which elements drive the depth shift. Fourth, \textbf{longitudinal studies} examining whether the depth effect persists across many sessions would test whether the effect is robust or diminishes with repeated exposure. Finally, investigating \textbf{alternative fear-based formulations}---beyond the PUA style tested here---would determine whether all fear-based approaches are equally ineffective or whether specific threat formulations might produce different results.

\section{Conclusion}

Trust-based motivational framing in system prompts significantly changes how AI agents investigate problems. In Study~1, the trust-framed agent found 15\% fewer surface-level issues but 59\% more hidden issues ($p = 0.002$, Cohen's $d = 2.28$), investigated 83\% more thoroughly ($p = 0.002$, Cohen's $d = 3.51$), and went beyond the stated task in every scenario. The behavioral signature is a shift from breadth-first surface scanning to depth-first investigation---quality over quantity.

In Study~2, we replicated the depth advantage across 5 independent runs and added a PUA fear-based condition, finding that fear-based prompting produced no significant improvement over baseline---trust produces depth, fear does not. NoPUA agents took 74\% more investigative steps ($p = 0.008$) and found 25\% more hidden issues ($p = 0.016$) than baseline, while PUA agents were statistically indistinguishable from baseline on all metrics (all $p > 0.3$).

These findings carry practical implications for prompt engineering: the \emph{how you frame} matters as much as the \emph{what you ask}---but only certain framings matter. Fear is motivationally inert for AI agents; trust, paired with structured methodology, produces genuine behavioral change. As AI agents take on increasingly autonomous roles in software development, the design of their motivational framing deserves the same attention as their technical capabilities. The NoPUA methodology, experimental data, and benchmark are available at \url{https://github.com/wuji-labs/nopua}.

\section*{Acknowledgments}

We thank Mr.\ Zhongshu, Mr.\ Zhangyu, and Mr.\ Pan for their guidance, friendship, and trust---the very qualities this paper argues for. We also thank the NoPUA community for valuable feedback and discussion.

\end{document}